\documentclass[prb,showpacs, preprint]{revtex4}%
\usepackage{amsfonts}
\usepackage{amsmath}
\usepackage{amssymb}
\usepackage[dvips]{graphicx}%
\usepackage{float}
\usepackage{epsfig}
\usepackage{subfigure}

\usepackage{enumitem}

\begin{document}
\title{Stabilizing intrinsic defects in SnO$_{2}$}%

\author{Gul Rahman}
\email{gulrahman@qau.edu.pk}

\affiliation{Department of Physics,
Quaid-i-Azam University, Islamabad 45320, Pakistan}
\author{Naseem Ud Din }
\affiliation{Department of Physics,
Quaid-i-Azam University, Islamabad 45320, Pakistan}

\author{V\'{\i}ctor M. Garc\'{\i}a-Su\'arez}
\affiliation{Departamento de F\'isica, Universidad de Oviedo, 33007 Oviedo, Spain}
\affiliation{Nanomaterials and Nanotechnology Research Center (CINN, CSIC), Spain}
\affiliation{Department of Physics, Lancaster University, Lancaster LA1 4YW, United Kingdom}
\author{Erjun Kan}

 \affiliation{Key Laboratory of Soft Chemistry and Functional Materials (Ministry of Education), and Department of Applied Physics, Nanjing University of Science and Technology, Nanjing, Jiangsu 210094, P. R. China.}


\begin{abstract}
The magnetism and electronic structure of Li-doped SnO$_{2}$ are investigated using first-principles LDA/LDA$+U$ calculations. We find that Li induces magnetism in SnO$_{2}$ when doped at the Sn site but becomes non-magnetic when doped at the O and interstitial sites. The calculated formation energies show that Li prefers the Sn site as compared with the O site, in agreement with previous experimental works. The interaction of Li with native defects (Sn V$_\mathrm{Sn}$ and O V$_\mathrm{O}$ vacancies) is also studied, and we find that Li not only behaves as a spin polarizer, but also a vacancy stabilizer, i.e. Li significantly reduces the defect formation energies of the native defects and helps the stabilization of magnetic oxygen vacancies. The electronic densities of states reveals that these systems, where the Fermi level touches the conduction (valence) band, are non-magnetic (magnetic).
\end{abstract}


\maketitle
\
\section{Introduction}

Diluted magnetic semiconductors (DMS) have recently been a major focus of magnetic semiconductor research. In the recent past, ferromagnetism (FM) above room temperature has been extensively investigated in transition metal (TM)-doped wide band gap oxide semiconductors, e.g. ZnO~\cite{ZnO,ZO} and TiO$_2$. ~\cite{tio2,tio3,tio4} Among these wide band gap semiconductors SnO$_{2}$ has long been a material of interest for applied and pure research purposes. In 2002, Ogale \textit{et al.}~\cite{ogale} not only reported room temperature FM in Co-doped SnO$_2$, but also very large magnetic moments. Later on, density functional theory (DFT) calculations showed that Sn vacancies, V$_\mathrm{Sn}$, can induce magnetism in SnO$_2$ and the observed large magnetic moment was attributed to its influence.~\cite{gu} Recently, several experimental and theoretical reports showed that cation vacancies can induce magnetism not only in SnO$_2$, but also in ZnO, TiO$_2$, ZrO$_2$, In$_2$O$_3$,CeO$_2$ HfO$_2$.~\cite{01, 02, 03, 04, 05, 06}

Following the prediction of magnetism without TM impurities, much interest was also diverted to magnetism induced by light element doping in host semiconductor matrices, e.g., K, N, Mg, and C-doped SnO$_2$. ~\cite{KCa,Nit, Mg,Rahman2010,Hong} There are experimental evidences which clearly demonstrate room temperature FM induced by light elements.\cite{cexp, ZnOprl} However, the exact nature of magnetism in these semiconductor oxides is still under debate. Recent experimental reports claim that pristine SnO$_{2}$ nanocrystalline thin films exhibit room temperature FM, but the magnetic moment is suppressed when doped with Gd.\cite{S.Ghosh} The presence of large amounts of singly ionized oxygen vacancies (V$_\mathrm{O}^{+}$), rather than V$_\mathrm{Sn}$ or V$_\mathrm{O}$ were found to be responsible for the observed FM in pristine SnO$_{2}$ thin films.\cite{S.Ghosh} On the other hand Chang \textit{et al.}\cite{Chang} observed FM in very thin pristine SnO$_{2}$ films, which was induced by the presence of oxygen vacancies located near the film surface. There is also a belief that the observed FM in undoped SnO$_{2}$ mainly originates from bulk double oxygen vacancies.\cite{Vgolo} In contrast, Wang \textit{et al.}~\cite{Wang} studied nanosheets of SnO$_{2}$ and they found that the saturation magnetization of all their annealed samples did not feature mono-dependence on oxygen vacancies, whereas an Sn vacancy related origin was accounted for variations in the magnetization of their studied samples.

Light elements can induce magnetism, but the observed magnetism has also a linkage with native defects. However, whether magnetism is induced by cation or anion vacancies in the pristine host material or doped with light elements, the defect formation  energies of native defects, which are important for magnetism, are very high. Oxygen vacancies have lower formation energies,\cite{alex} but neutral V$_\mathrm{O}$ does not induce magnetism in oxides.\cite{gu,stefano,ZnOprl} The major issue in magnetism induced by vacancies is therefore that these magnetic systems have high formation energies. ~\cite{Zn, chro} Consequently, to realize defects-driven magnetism experimentally it is essential to reduce the defect formation energy of the host material (SnO$_2$ in our case). We choose Li as a dopant~\cite{ptype} that can modify the defect formation energies of native defects in SnO$_2$. We show that Li can indeed significantly reduce the formation energies of various types of defects. This is, as far as we know, the first theoretical study performed on the interaction of Li with native defects in SnO$_2$.

\section{Computational Models and Method}

SnO$_{2}$ is known to crystallize in the rutile structure with space group $P4_2/mnm$ or $D^{4h}_{14}$ (SG136) under ambient conditions.~\cite{group} We used our previously optimized lattice parameters.~\cite{gu} To simulate the defects a $2\times2\times2$ supercell of SnO$_2$ containing 48 atoms was employed. We considered the following models to investigate the detailed energetics and magnetic properties of different defects in SnO$_2$.

\begin{figure}[b]
\begin{center}
\includegraphics[width=1.0\textwidth]{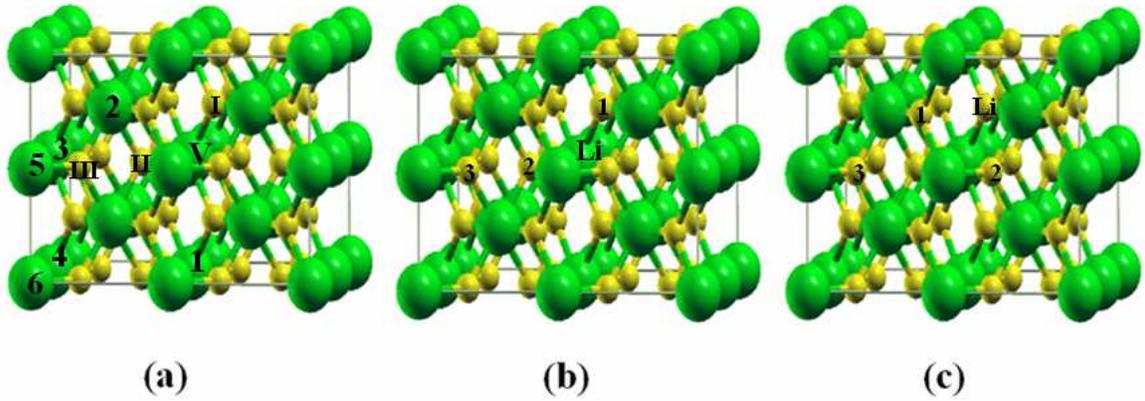}
\caption{(color online) A $2\times 2 \times 2$ supercell of SnO$_2$. In (a) V represents the Sn vacancy, which is fixed at the center of the supercell. The positions of Li doped at different Sn sites are marked as 1-6, whereas the positions of Li doped at the O sites are represented by I,II, and III. In model (b) the central atom represents the Li atom doped at the Sn site and the O vacancies are marked as 1-4. In model (c) Li is doped at the O site, marked as Li, and the oxygen vacancies are created at the oxygen sites marked as 1, 2, and 3. Big (green) and small (blue) balls represent Sn and O atoms, respectively.}
\label{z}
\end{center}
\end{figure}

\begin{enumerate}[label=\alph*)]
\item Li-doped SnO$_{2}$: Li was doped at the Sn, O, and
interstitial sites, denoted as Li$_\mathrm{Sn}$, Li$_\mathrm{O}$, and Li$_\mathrm{int}$, respectively.

\item V$_\mathrm{Sn}$+ Li$_{\mathrm{Sn}j}$:In this case,
V$_\mathrm{Sn}$ was fixed at the center of the supercell, and Li was doped at different Sn ${j}$ sites, with $j=1-6$ (see Fig.~\ref{z}(a)).

\item V$_\mathrm{Sn}$+ Li$_{\mathrm{O}j}$: In this case,
V$_\mathrm{Sn}$ was fixed at the center of the supercell, and Li was doped at different O ${j}$ sites, with $j$=I, II, III (see Fig.~\ref{z}(a)).

\item V$_\mathrm{Sn}$+ Li$_\mathrm{int}$: Here V$_\mathrm{Sn}$ was
fixed at the center of the supercell and  Li was doped at interstitial sites.

\item V$_{\mathrm{O}j}$+ Li$_\mathrm{Sn}$: Here Li was doped at the
central Sn site and V$_\mathrm{O}$ was created at different O $j$ sites, with $j=1-4$ (see Fig.~\ref{z}(b))

\item V$_\mathrm{O}$+ Li$_{\mathrm{O}j}$: In this case,
V$_\mathrm{O}$ was fixed and Li was doped at different O $j$ sites, with $j=1,2,3$ (see Fig.~\ref{z}(c))

\item V$_\mathrm{O}$+ Li$_\mathrm{int}$: Here V$_\mathrm{O}$ was
fixed at the center of the supercell and Li was doped at interstitial sites.
\end{enumerate}

We performed calculations in the framework of density functional theory (DFT) ~\cite{DFT} using linear combination of atomic orbitals (LCAO) as implemented in the SIESTA code. ~\cite{siesta} We used a double-$\zeta$ polarized (DZP) basis set for all atoms, which included $s$, $p$ and $d$ orbitals in Sn and O (we polarized $p$ orbitals, which added additional 5 $d$ orbitals) and $s$ and $p$ orbitals in Li (we polarized an $s$ orbital, which added 3 $p$ orbitals). The local density approximation (LDA)~\cite{lda} was adopted for describing exchange-correlation interactions. We used standard norm-conserving pseudopotentials ~\cite{ps} in their fully nonlocal form. ~\cite{pss} Atomic positions and lattice parameters were optimized, using a conjugate-gradient algorithm, ~\cite{cg} until the residual Hellmann-Feynman forces converged to less than 0.05 eV/\AA. A cutoff energy of 400 Ry for the real-space grid was adopted. This energy cutoff defines the energy of the most energetic plane wave that could be represented on such grid, i.e. the larger the cutoff the smaller the separation between points in the grid ($E\sim G^2\sim 1/d^2$, where $\vec G$ is a reciprocal vector and $d$ is the separation between points). Using the relaxed LDA atomic volume/coordinates, we also carried out LDA$+U$ calculations by considering the on-site Coulomb correction ($U= 6.0$ eV) between
the $p$-orbital electrons of O.\cite{ldau1,ldau2} We also cross checked some calculations with the VASP code, which uses plane waves and ultrasoft pseudopotentials.\cite{VASPc}

The concentration of dopants and vacancies in a crystal depends upon its formation energies ($E_\mathrm{f}$). The chemical potentials, which vary between the stoichiometric and diluted limits, rely on the material growth conditions and boundary conditions.
The chemical potentials of Sn and O depend on whether SnO$_2$ is grown under
O-rich or Sn-rich growth conditions.
We have calculated the formation energies under equilibrium conditions, O-rich conditions and Sn-rich conditions. Under equilibrium conditions, for SnO$_2$ the chemical potentials of O and Sn satisfy the relationship $\mathrm{\mu_{Sn} + 2\mu_O = \mu_{SnO_{2}}} $, where $\mu_\mathrm{SnO_{2}}$ is the chemical potential of bulk SnO$_2$ is a constant value calculated as the total energy per SnO$_2$ unit formula ~\cite{form}, and $\mu_\mathrm{O_{2}}$ is the chemical potential of O which is calculated as total energy per atom of O$_{2}$ molecule . The chemical potential of Sn $\mu_\mathrm{Sn}$ is calculated as $\mu_\mathrm{Sn}$=E(Sn$^\mathrm{metal}$), where =E(Sn$^\mathrm{metal}$) is the total energy per atom of bulk Sn.  For equilibrium condition we used $\mu_\mathrm{Sn}=\mu_\mathrm{SnO_{2}}-2\mu_\mathrm{O}$ and $\mu_\mathrm{O}=\left(\frac{\mu_\mathrm{SnO_{2}}-\mu_\mathrm{Sn}}{2} \right)$. Under Sn rich conditions, $\mu_\mathrm{Sn}$=E(Sn$^\mathrm{metal}$), $\mu_\mathrm{O}=\left(\frac{\mu_\mathrm{SnO_{2}}-\mu_\mathrm{Sn}}{2} \right)$. O rich condition gives $\mu_\mathrm{O}$ = $\left(\frac{1}{2}\right)$E(O$_2$), $\mu_\mathrm{Sn}=\mu_\mathrm{SnO_{2}}-2\mu_\mathrm{O}$. Note that
$\mu_\mathrm{Sn}$ and $\mu_\mathrm{O}$ are not independent, but vary between
the Sn-rich and O-rich limits under a constraint defined by the equilibrium
condition of SnO$_{2}$. The Sn-rich limit corresponds to
the upper limit of $\mu_\mathrm{Sn}$ and also the lower limit of $\mu_\mathrm{O}$. Therefore it is expected that different chemical potential will give different defect formation energies.\cite{defects-2012}

The formation energies for systems with intrinsic defects, either Sn or O vacancies, can be calculated in the following way:

\begin{equation}
E_\mathrm{f}=E^\mathrm{d}-E^\mathrm{p}+n\mu_\mathrm{X},
\end{equation}

\noindent where $\mu_\mathrm{X}$ is chemical potential of X (X = Sn, O), $n$ is the number of atoms removed from the system and $E{^\mathrm{d}}$ and $E^\mathrm{p}$ are the total energies of the defected and the pure system, respectively.

The formation energy of the Li doped system can be calculated as

\begin{equation}
E_\mathrm{f}=\frac{1}{n}\left(E^\mathrm{d}-E^\mathrm{p}+n\mu_\mathrm{X} - m\mu_\mathrm{Li}\right),
\end{equation}

\noindent where $\mu_\mathrm{Li}$ is the chemical potential of bulk lithium, calculated as the total the energy per unit cell of bulk Li, $n$ is number of atoms removed and $m$ is number of atoms added to the system.
We note again that some of these energies, which are very sensible, were cross checked with the VASP code.

\section{Results and Discussions}

As stated above, we relaxed all the systems, so we will only discuss the relaxed data. Comparisons will be made with the unrelaxed data where necessary. It is important to discuss the implication of the LDA$+U$ approach before starting the discussion of the calculated results. Our LDA calculations show that SnO$_{2}$ has a band gap 0.9 eV which is comparable with previous LDA calculations.\cite{PRB2012} It is known that LDA underestimates the band  gap of materials, which can be corrected by applying the LDA$+U$ approcah. The LDA$+U$ calculated band gap of SnO$_{2}$ in our case is $\sim 3.10 $ eV which is comparable with the experimental and theoretical values of 3.20 eV.\cite{iop,dean} It is to be noted that the formation energies of relaxed LDA systems can also be affected by the $U$ term, which will be discussed wherever required. For electronic structures (densities of states), we will only show the LDA$+U$ calculated results, because LDA does not reproduce accurately the band gap, as stated above.

First, we will focus on Li-doped SnO$_{2}$ and then we will move to discuss the interaction of Sn and O vacancies with Li.

\subsection{Li-doped SnO$_2$}

First, we calculate, using the above equations, the defect formation energies of the doped SnO$_2$ systems, where Li was doped at the Sn, O, and interstitial sites, as stated before. The calculations (see Table (\ref{t1})) show that the doping of Li at the Sn site has the lowest formation energy (-1.03 eV) under stoichiometric and O-rich conditions, as compared with the O site. Although Li at the interstitial site has the lowest formation energy among the three dopant sites it is not important from the magnetism point of view. For comparison purposes, the formation energies of the LDA$+U$ systems are also given in Table (\ref{t1}). Our detailed structural relaxation analysis shows that structural relaxation is not very effective when Li is doped at the Sn site. This behavior can easily be understood in terms of the atomic sizes of the Sn and Li atoms. The atomic radii of Sn and Li are 1.41 \AA and 1.45 \AA, respectively. The doped Li is surrounded by  six O atoms and the optimized  bond lengths between them (LiO distances) have two different values, 1.97~\AA, and 2.03~\AA, which are comparable to the pure bond lengths of SnO in SnO$_2$, i.e., 2.04~\AA~and  2.08\AA. Due to the radius similarity, Li at the Sn site will not distort the structure and it is therefore expected that the Sn site will be favorable for Li doping. Indeed our calculations also show that Li at the Sn site has the lowest formation energy as compared with the O site, which agrees with experimental observations.~\cite{ptype} On the other hand we found that the unrelaxed Li at the O site has a large formation energy under the equilibrium condition. The formation energy of Li at the O site was very large when structural relaxation was not allowed but decreased significantly when structural relaxation was permitted, which produced a large structural distortion. The structural analysis showed that when Li is doped at the O site, it goes to an interstitial site and leaves behind an O vacancy. So, this indicates that the formation energy of O is lowered mainly due to a structural deformation which is accompanied by the movement of Li to an interstitial position. At the same time, the system remains nonmagnetic. To be more confident, we also carried out separate calculations on Li at interstitial sites with and without O vacancies, and we got the same conclusion, which will be discussed below.
Comparing the LDA and LDA$+U$ calculated formation energies, it is clear that LDA$+U$ changes only the formation energy of the Li$_{\rm O}$ system, whereas the rest of the cases remain similar.

\begin{figure}
\centering
\begin{tabular}{ccccccc}
\epsfig{file=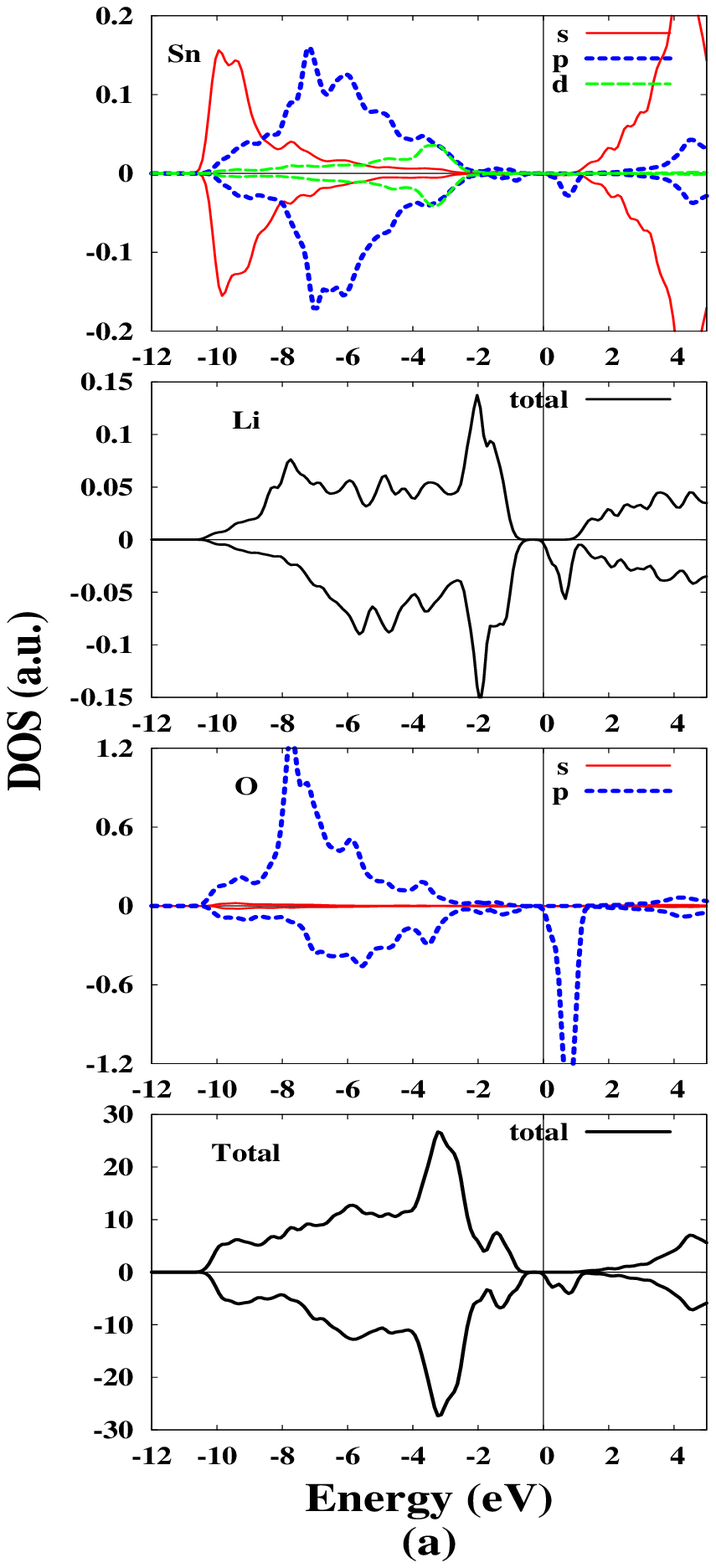,width=0.36\linewidth,height=1.2\linewidth,clip=} &
\epsfig{file=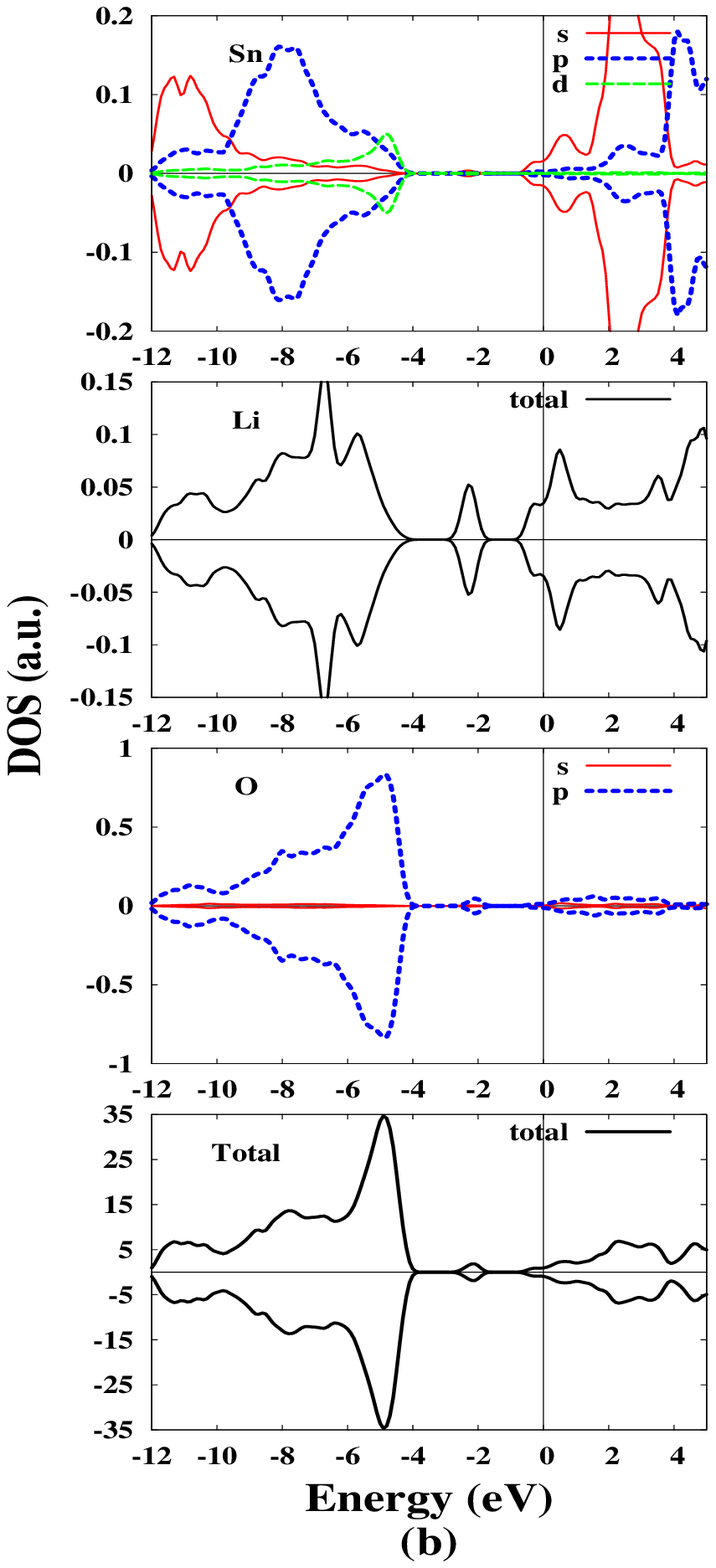,width=0.36\linewidth,height=1.2\linewidth,clip=} &
\epsfig{file=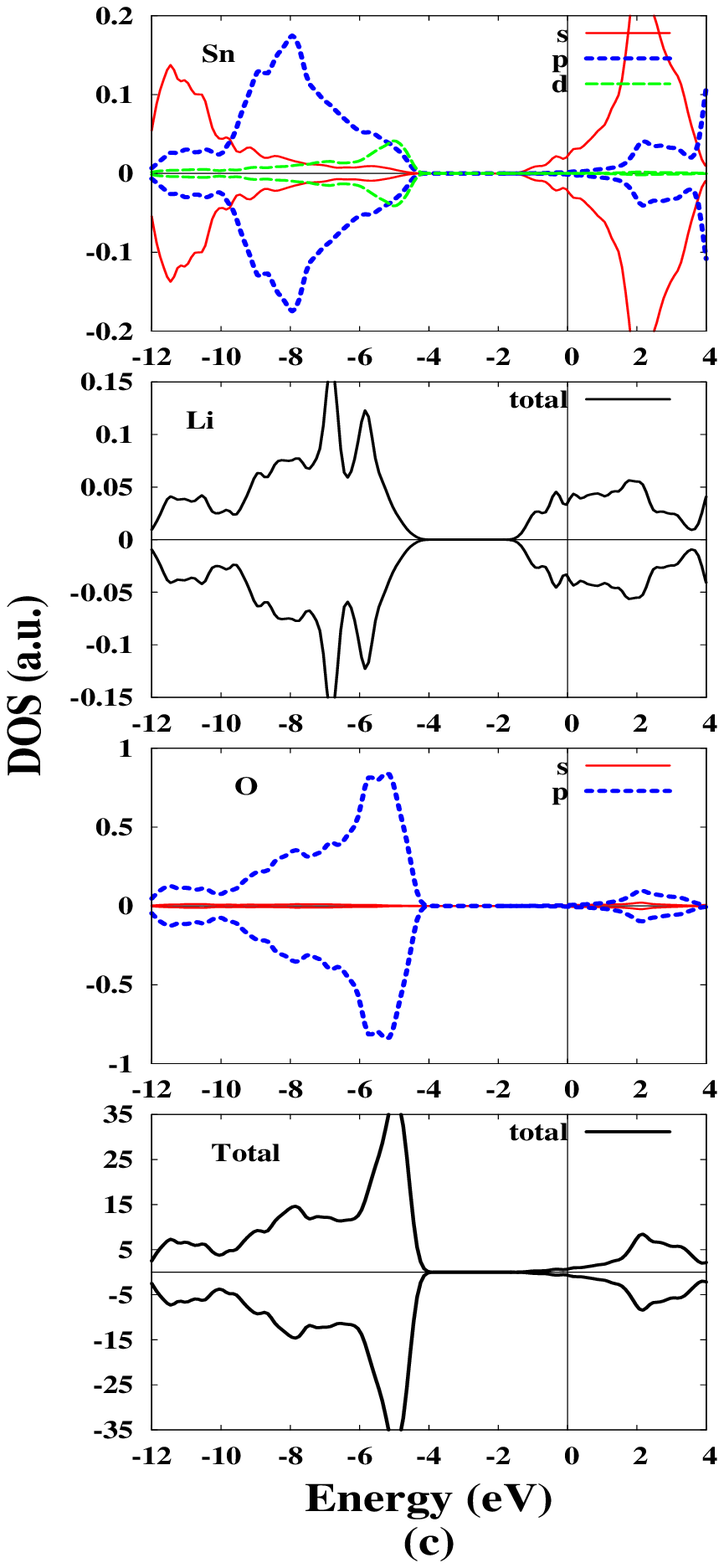,width=0.36\linewidth,height=1.2\linewidth,clip=}
\end{tabular}
\caption[Optional caption for list of figures]{(color online) The LDA$+U$ calculated total and partial density of states (DOS) of (a) Li$_\mathrm{Sn}$, (b) Li$_\mathrm{O}$, and (c) Li$_\mathrm{int}$ systems. Solid (red) and dashed (blue, green) lines represent $s$ and $p$, $d$  states, respectively. In the bottom panels the solid line represents the total DOS. The positive (negative) DOS shows majority (minority) spin states. The Fermi level ($E_\mathrm{F}$) is set to zero.}
\label{dos1}
\end{figure}

As stated above, doped Li prefers the Sn site under stoichiometric and O-rich conditions, where it has a large magnetic moment (3.00 $\mu_\mathrm{B}$). Since the nominal valence of Sn in a perfect SnO$_2$ is Sn$^{4+}$, and Li is a cation with valence of Li$^{1+}$,~\cite{ptype} when Li is doped at the Sn site in pure SnO$_2$, it donates one electron to compensate one hole among the four holes generated by the Sn deficiency. The three uncompensated holes, localized at the O sites, give a magnetic moment of 3.00 $\mu_\mathrm{B}$ per supercell. Table (\ref{t1}) clearly shows that LDA$+U$ does not change the total magnetic moments of the doped/defected systems, in agreement with previous theoretical calculations which have also shown that the inclusion of the $U$ term does not change the magnetic moments caused by vacancies.~\cite{Sanvito,Fernandes} To see the atomic origin of magnetism in Li-doped SnO$_2$, we calculated the total and atom projected partial density of states(PDOS) of the Sn, O, and Li atoms (see Fig.~\ref{dos1}). As can be seen, the substitutional Li impurity polarizes the host band, which gives rise to a induced impurity peak close to the valance band edge in the spin-down part of the total DOS, while the spin-up part is influenced slightly. Such a behavior suggests that Li behaves as a $p$ type dopant in SnO$_{2}$.~\cite{ptype}
The low lying $s$ orbitals of Li are spin-polarized and strongly hybridized with the $p$
orbitals of O. The Fermi energy is mainly dominated by the $p$ orbitals of O, which indicates that magnetism is mainly induced by the $p$ orbitals and localized
at the O atom. Indeed, the oxygen atoms surrounding the doped Li are the main ones that contribute to magnetism.  We also see that the spin down state is partially occupied which contributes to the magnetic moment. The majority $s$ spin states of Li are completely occupied and the minority spin states are partially occupied, leading to a significant spin splitting near the Fermi level.

\begin{figure}[b]
\begin{center}
\includegraphics[width=0.8\textwidth]{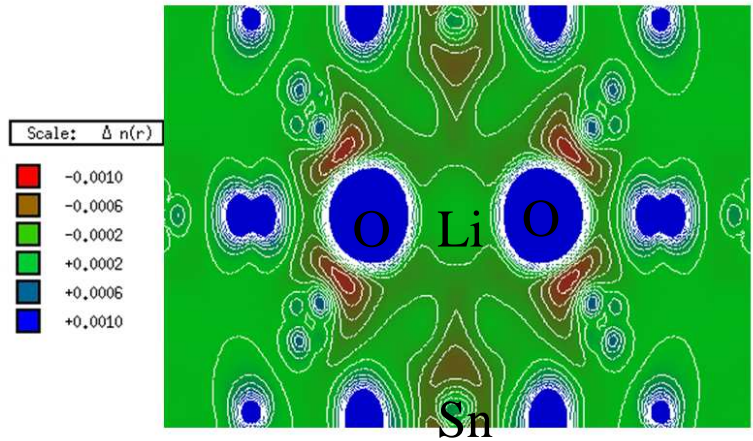}
\caption{(color online) Spin density contours for the Li-doped SnO$_2$ (110) plane. Labels show atomic sites.}
\label{rho}
\end{center}
\end{figure}

To further emphasize on the nature of magnetism, the spin density contours are shown in Fig.~\ref{rho}. It is interesting to see that Li polarizes the O and Sn atoms in opposite directions, and the polarization is not only limited to the nearest O atoms surrounding the Li atoms, but it also spreads to other O atoms which are far from the Li.
It seems that the origin of magnetism in light elements doped-SnO$_2$ or Sn vacancies in SnO$_2$ is the same i.e., the polarization of the surrounding O atoms. However, the remarkable feature of Li doped SnO$_2$ is that there is a very small (negligible) induced magnetic moment at the Li site. This behavior is quite different from other light elements doped SnO$_{2}$ systems,~\cite{KCa,Nit,Mg} which suggests that Li behaves as a spin polarizer in SnO$_2$. We note that Li doped at the O and interstitial sites does not induce magnetism (see Fig.~\ref{dos1}).
Test calculations on a supercell of  $2\times2\times3$ (containing 72 atoms) were also carried out and we found the same behavior, i.e. Li at the Sn site gives 3.00 $\mu_\mathrm{B}$ magnetic moment per supercell, whereas Li at the O site does not show any magnetic moment. Using a larger supercell does change the defect formation energy of the defected systems.\cite{Rahman2013,juliana2013} Additional calculations were also carried  out to see the magnetic coupling between the Li atoms, since it is known that the magnetic coupling between the impurity atoms depend on the separation between the defects/impurities.\cite{gu,znoprl-2005} The results show that Li couples ferromagnetically.~\cite{SnO2Li-2013}

\subsection{Interaction of Sn vacancies with Li}

We will move now to discuss the interaction of intrinsic defects (Sn and O vacancies) with Li. We found that a single Sn vacancy induces a very large magnetic moment (LDA and LDA$+U$)$\sim$4.00 $\mu_\mathrm{B}$, which agrees with previous calculations.~\cite{gu} However, V$_\mathrm{Sn}$ has a very large formation energy in both O-rich (6.87 eV) and Sn-rich(14.44 eV) conditions, which is not significantly reduced after structural relaxation. To decrease the formation energy of V$_\mathrm{Sn}$, which is the core of this article, we doped Li at different Sn sites (see Figure~\ref{z}), with the Sn vacancy at the center of the supercell. The formation energies are given in Table (\ref{t2}). As can be seen, by  doping Li at the Sn site, the defect formation energy of V$_\mathrm{Sn}$ is significantly reduced. The case V$_\mathrm{Sn}$+Li$_\mathrm{Sn3}$ has the lowest formation energy $\sim$1.54 eV, which shows that Li doping is much more favorable energetically in SnO$_2$ than previously reported Zn~\cite{Zn} and Cr~\cite{chro} doping: The formation energies of the system with Zn$_\mathrm{Sn}$+V$_\mathrm{Sn}$ are 16.50 eV and 7.00 eV under O-poor and the O-rich conditions,~\cite{Zn} repectively. Doping Cr in SnO$_2$ with V$_\mathrm{Sn}$ produces formation energies $\sim $12.00 eV and 2.00 eV under O-poor and O-rich conditions,~\cite{chro} respectively. All these previous calculations were carried out without U. In our case the formation energies are 9.10 eV and 1.54 eV under O-poor and O-rich conditions, respectively. It is interesting to see that Li not only stabilizes V$_\mathrm{Sn}$, but produces also the largest magnetic moment, 7.00 $\mu_\mathrm{B}$, per supercell. The oring of such magnetic moment is due to the three uncompensated holes of Li at the Sn site along with the four holes of neutral V$_\mathrm{Sn}$.

\begin{figure}[b]
\begin{center}
\includegraphics[width=1.0\textwidth]{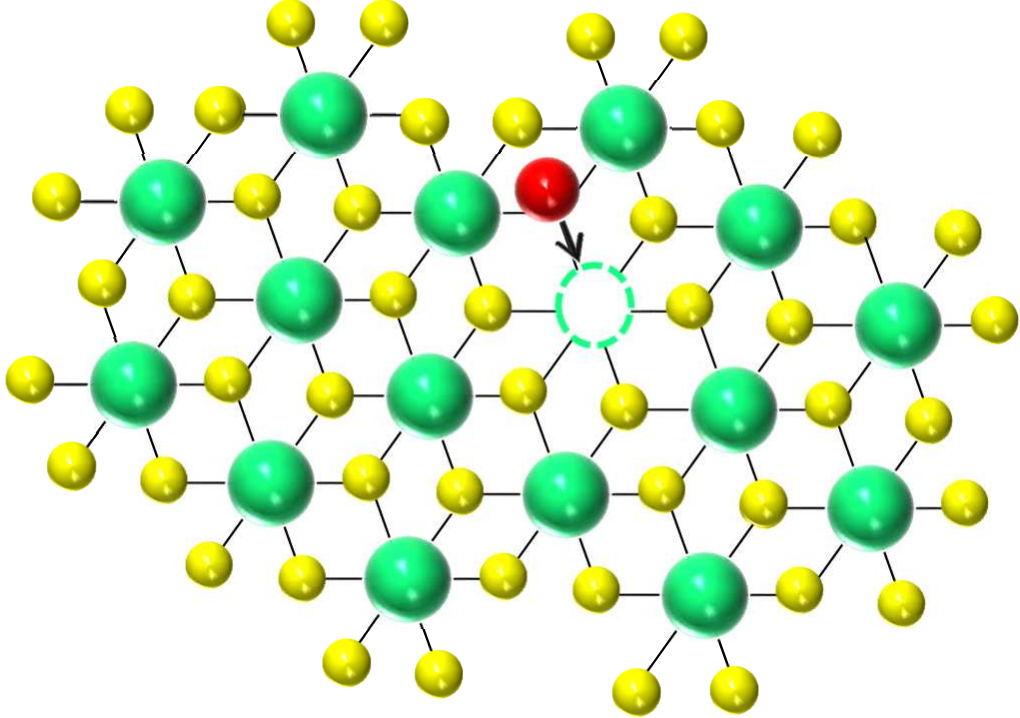}
\caption{(color online) Schematic representation of the structural relaxation in the Li$_\mathrm{O}+V_\mathrm{Sn}$ system. Green, yellow, and red balls represent Sn, O, and Li atoms, respectively. The Sn vacancy (V$_\mathrm{Sn}$) is represented by a green dashed circle. The arrow represents the movement of the Li atom towards V$_\mathrm{Sn}$ during the structural relaxation.}
\label{stable}
\end{center}
\end{figure}

Once found that Li at Sn sites significantly reduces the defect formation energy of Sn vacancies we doped Li at the O site in the presence of  V$_\mathrm{Sn}$. In this case, we also fixed V$_\mathrm{Sn}$ and doped Li at different O sites, mainly considering those O atoms which are near and far from V$_\mathrm{Sn}$. The calculated formation energies are shown in  Table (\ref{t2}). We can see that the unrelaxed systems have large positive formation energies (endothermic processes), but after structural relaxation they turn to negative formation energies (exothermic processes). This effect of structural relaxation on the defect formation energies was also confirmed using VASP. For example, the relaxed (unrelaxed) defect formation energy of the V$_\mathrm{Sn}$+Li$_\mathrm{O_{II}}$ system under equilibrium conditions was found to be -3.74 (5.39) eV using the SIESTA code, whereas VASP gave -4.08 (4.51) eV. When Li was doped at O sites around V$_\mathrm{Sn}$ it moves to an interstitial configuration (near V$_\mathrm{Sn}$) after structural relaxation. This exothermic process leaves behind an oxygen vacancy and an interstitial Li near V$_\mathrm{Sn}$ (see Fig.~\ref{stable}), which indicates that doped Li does not prefer the O site, but the interstitial site.
As shown also above, the Li$_\mathrm{O}$ and Li$_\mathrm{int}$ systems have no magnetic moment, whereas V$_\mathrm{Sn}$ has a large magnetic moment. However, when Li is doped at the O site, three out of the four holes generated by V$_\mathrm{Sn}$ are compensated by the 2$^{-}$ charge state of V$_\mathrm{O}$ and one electron of Li, leaving behind a single hole that gives a magnetic moment of 1.00 $\mu_\mathrm{B}$. When Li is far from V$_\mathrm{Sn}$ (V$_\mathrm{Sn}$+Li$_{\mathrm{OIII}}$ case), it does not form a stable structure.

We discuss now the interaction of Sn vacancies with Li$_\mathrm{int}$.  In this case, we doped Li at an interstitial site which was 5.37 \AA~ away from V$_\mathrm{Sn}$. The defect formation energy of this system is given in Table (\ref{t2}). Although the Li$_\mathrm{int}$ + V$_\mathrm{Sn}$ system behaves endothermically, the formation energy of V$_\mathrm{Sn}$ decreases a lot as compared with the V$_\mathrm{Sn}$ system without Li$_\mathrm{int}$. Note that the structural relaxation does not change the sign of $E_\mathrm{f}$.
The Li$_\mathrm{int}$ + V$_\mathrm{Sn}$ system also shows a magnetic moment, 3.00 $\mu_\mathrm{B}$ per supercell. In other words, when Li is at the interstitial site it interacts  with V$_\mathrm{Sn}$ and reduces its magnetic moment.
From the  thermodynamics of different defects in SnO$_2$, we can conclude that among all these defects the V$_\mathrm{Sn}$ + Li$_\mathrm{O}$ complex has the lowest energy. In the presence of vacancies (Sn and O), Li prefers to occupy the O site rather than going into Sn or interstitial sites. Although, interstitial Li also reduces the formation energy of V$_\mathrm{Sn}$, it is not a favorable process. Note that different defective systems of Li-doped ZnO showed the same type of behavior in a previously reported work.~\cite{ZnOprl}

It is very interesting that irrespective of the location of Li in Sn defective systems, it always reduces the formation energy of V$_\mathrm{Sn}$ and the systems remains magnetic both with LDA and LDA$+U$. To further understand the electronic structure of doped SnO$_2$ and the interaction of Li with native defects, we show the LDA$+U$ calculated total and atom projected densities of states in Fig.~\ref{dos2}. The interaction of Li with native defects mainly affects the local DOS of the O atoms. Also, Li-O hybridizations can be seen around the Fermi levels. In all LDA cases, the oxygen spin-up $p$ orbitals were occupied whereas the spin-down $p$ orbitals were partially occupied, similar to the DOS of V$_\mathrm{Sn}$.\cite{gu} With LDA$+U$ the oxygen $p$ orbitals are almost totally occupied, as expected due to the $U$.


\subsection{Interaction of O vacancies with Li}

It is believed that Sn vacancies have higher formation energy as compared with O vacancies. Indeed, we found that the O vacancy has a defect formation energy (1.64 eV) much smaller than that for the Sn vacancy. We therefore considered also the interactions of O vacancies with Li. First, we doped Li at the Sn site and varied the position of the O vacancy. The calculated defect formation energies are shown in Table (\ref{t3}). Interestingly, doping of Li at the Sn site appreciably reduced the defect formation energy of V$_\mathrm{O}$. The defect complex V$_\mathrm{O3}$ + Li$_\mathrm{Sn}$ has the lowest formation energy, -3.69 eV, in equilibrium conditions. Note that our VASP calculations show -3.90 eV  for the same defect complex. Such a favorable process implies that oxygen vacancies also promote magnetism in doped systems. We must stress that oxygen vacancies by themselves are nonmagnetic,~\cite{gu} but however, become magnetic in the presence of  Li$_\mathrm{Sn}$, which is also magnetic. Either Li$_\mathrm{Sn}$ or V$_\mathrm{O}$ have large defect formation energies, but the defect complex V$_\mathrm{O}$ + Li$_\mathrm{Sn}$ has a lower energy than the individual defects. It is interesting to see that these defects complex always give magnetism with magnetic moments of 1.00 $\mu_\mathrm{B}$, no matter how far is V$_\mathrm{O}$ from Li$_\mathrm{Sn}$. Such behavior is different from Co-doped SnO$_{2}$ where it has been shown that V$_\mathrm{O}$ quenches magnetism when Co$_\mathrm{Sn}$ is away from V$_\mathrm{O}$.~\cite{refs-exp2} On the other hand, Ni-doped SnO$_{2}$ does not show magnetism without V$_\mathrm{O}$.~\cite{Nidoped} We therefore believe that Li-doped SnO$_{2}$ is another candidate material for spintronics. We also found that V$_\mathrm{O}$ and Li$_\mathrm{O}$ do not induce magnetism in SnO$_{2}$, as seen in Table (\ref{t3}), but the formation energy of this complex is always bigger than that of the V$_\mathrm{O}$ + Li$_\mathrm{Sn}$ complex, excluding under Sn-rich conditions. There are some experimental reports which claim that oxygen vacancies are responsible for room temperature ferromagnetism in oxides~\cite{refs-exp} but to date, none of them show that a single neutral  V$_\mathrm{O}$ is the main source of magnetism in SnO$_{2}$. Generally, it is believed that V$_\mathrm{O}$ promotes magnetism in doped SnO$_{2}$,\cite{refs-exp2,Nidoped} which is proven by our extensive calculations (see Table (\ref{t3})).
We also considered Li at different O sites in the presence of O vacancies but the calculated $E_\mathrm{f}$ show that these configurations are less stable than the V$_\mathrm{O}$ + Li$_\mathrm{Sn}$ (excluding Sn-rich conditions).

The interaction of O vacancies with Li$_\mathrm{int}$ was also taken into account. The calculated $E_\mathrm{f}$ of Li$_\mathrm{int}$+V$_\mathrm{O}$ are given in Table (\ref{t3}) which shows that the behavior of Li$_\mathrm{int}$+V$_\mathrm{O}$ is quite different from  Li$_\mathrm{int}$+V$_\mathrm{Sn}$. Li$_\mathrm{int}$ reduces the formation energy of V$_\mathrm{Sn}$, but however Li$_\mathrm{int}$ does not decrease the formation energy of  V$_\mathrm{O}$. Also such kinds of defect complexes have zero magnetic moments. Both LDA and LDA$+U$ calculations give the same magnetic moments, as found before. From Table (\ref{t2}) and Table (\ref{t3}), we can therefore see that $E_\mathrm{f}$ can be reduced by Li. On one hand, in the presence of a Sn vacancy the probable location of Li will be the O site, where structural relaxation is also expected. On the other hand, Li prefers the Sn site rather than the O site, when the interactions of O vacancies with Li$_{\mathrm{O}}$ is considered.

\begin{figure}
\centering
\begin{tabular}{ccccccc}
\epsfig{file=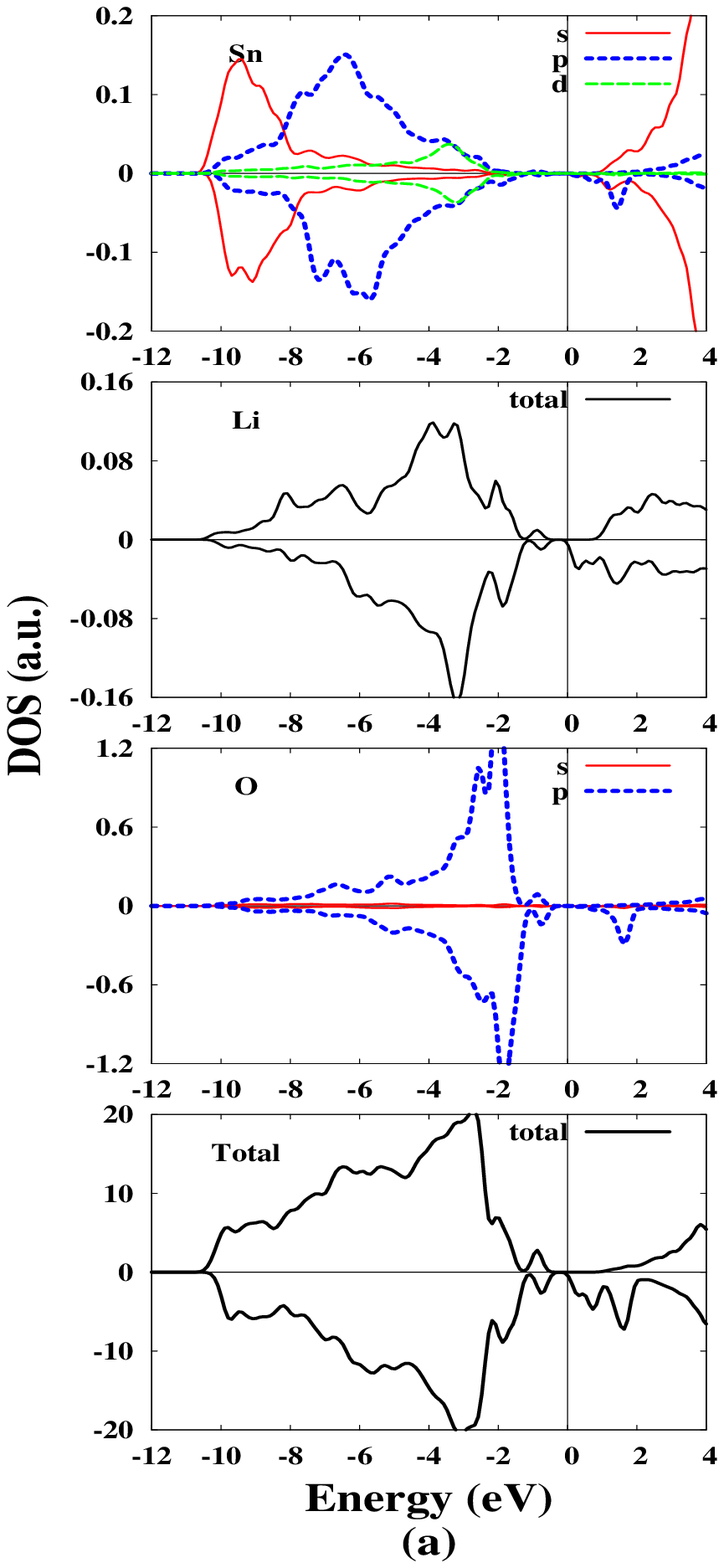,width=0.36\linewidth,height=1.2\linewidth,clip=} &
\epsfig{file=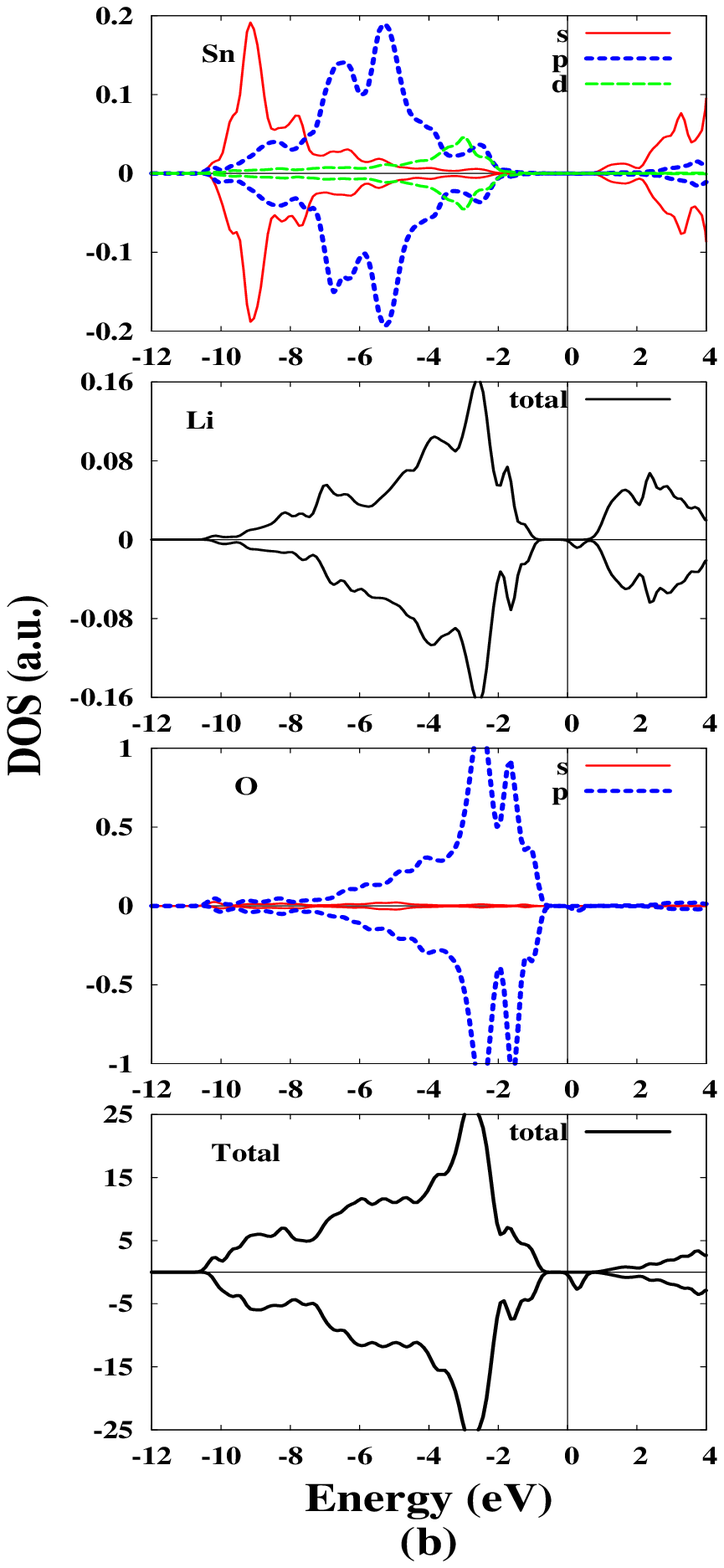,width=0.36\linewidth,height=1.2\linewidth,clip=} &
\epsfig{file=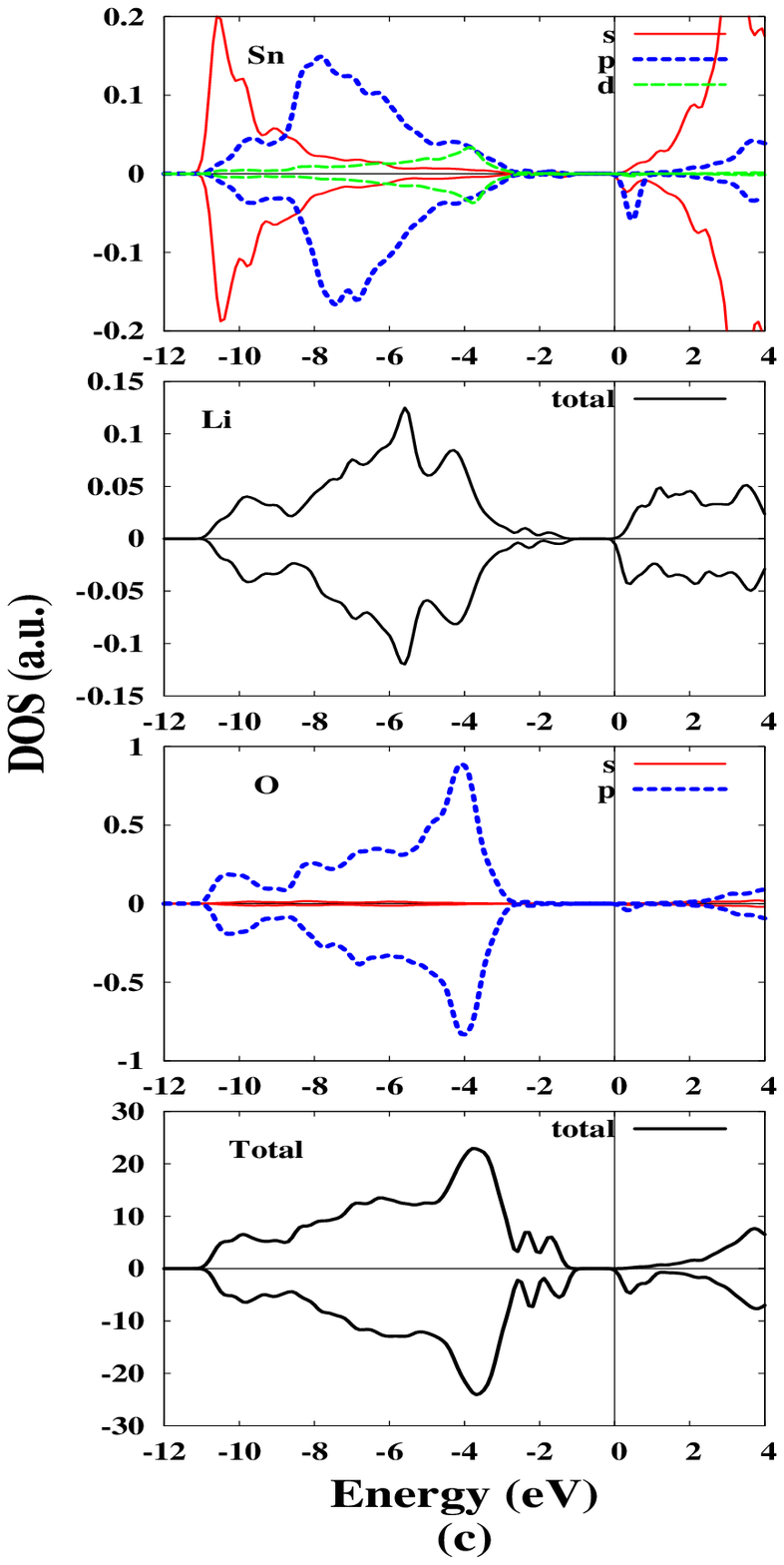,width=0.36\linewidth,height=1.2\linewidth,clip=}
\end{tabular}
\caption[]{(color online) The LDA$+U$ calculated total and partial density of states (DOS) of (a) V$_\mathrm{Sn}$+Li$_\mathrm{Sn}$, (b) V$_\mathrm{Sn}$+ Li$_\mathrm{O}$, and (c) V$_\mathrm{Sn}$+Li$_\mathrm{int}$ systems. Solid (red) and dashed (blue, green) lines represent $s$, $p$ and $d$  states, respectively. The bottom panel solid line represents the total DOS. The positive (negative) DOS shows majority (minority) spin states. The Fermi level ($E_\mathrm{F}$) is set to zero.}
\label{dos2}
\end{figure}

\begin{figure}
\centering
\begin{tabular}{ccccccc}
\epsfig{file=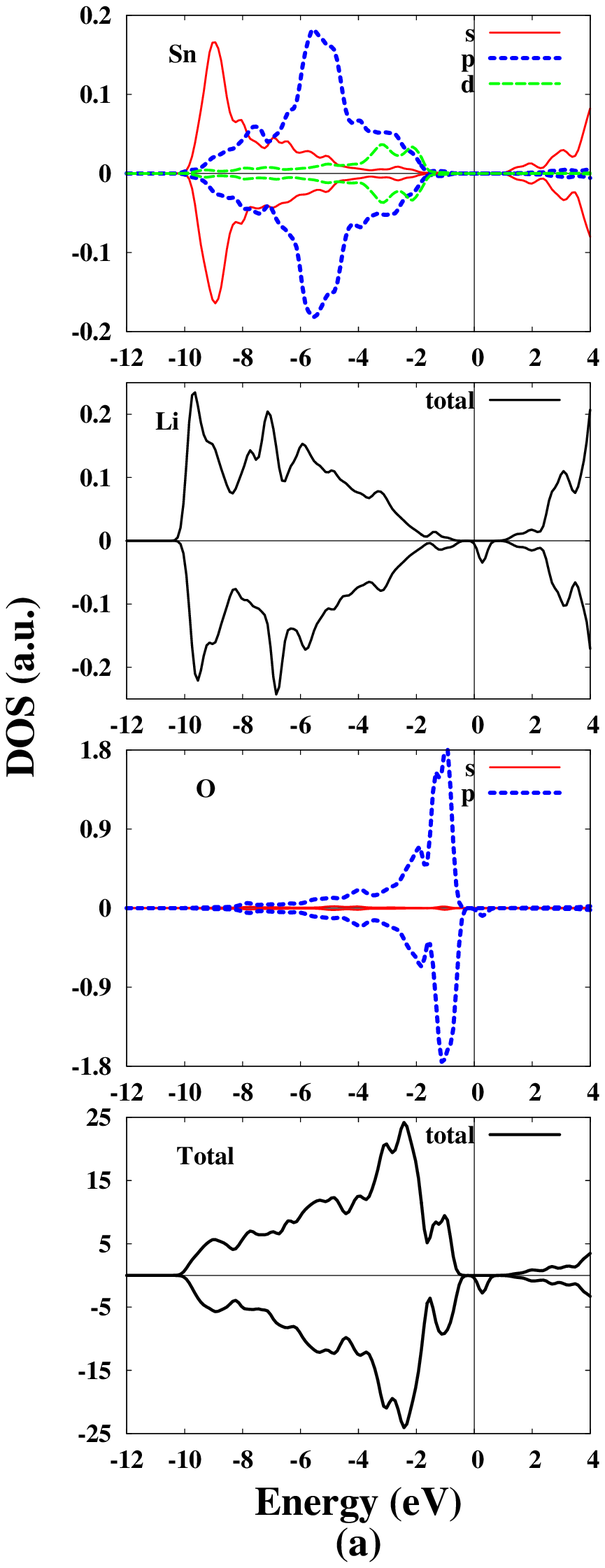,width=0.36\linewidth,height=1.2\linewidth,clip=} &
\epsfig{file=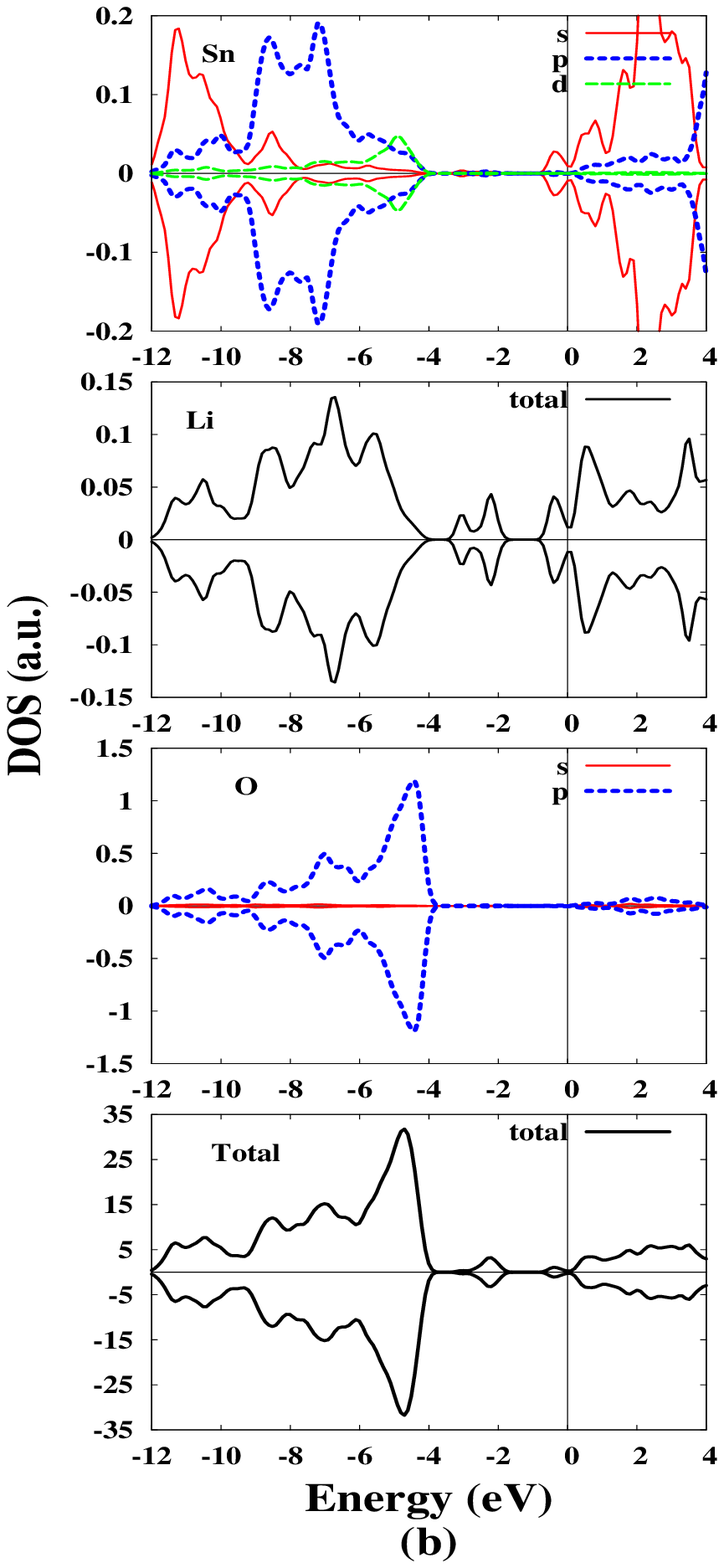,width=0.36\linewidth,height=1.2\linewidth,clip=} &
\epsfig{file=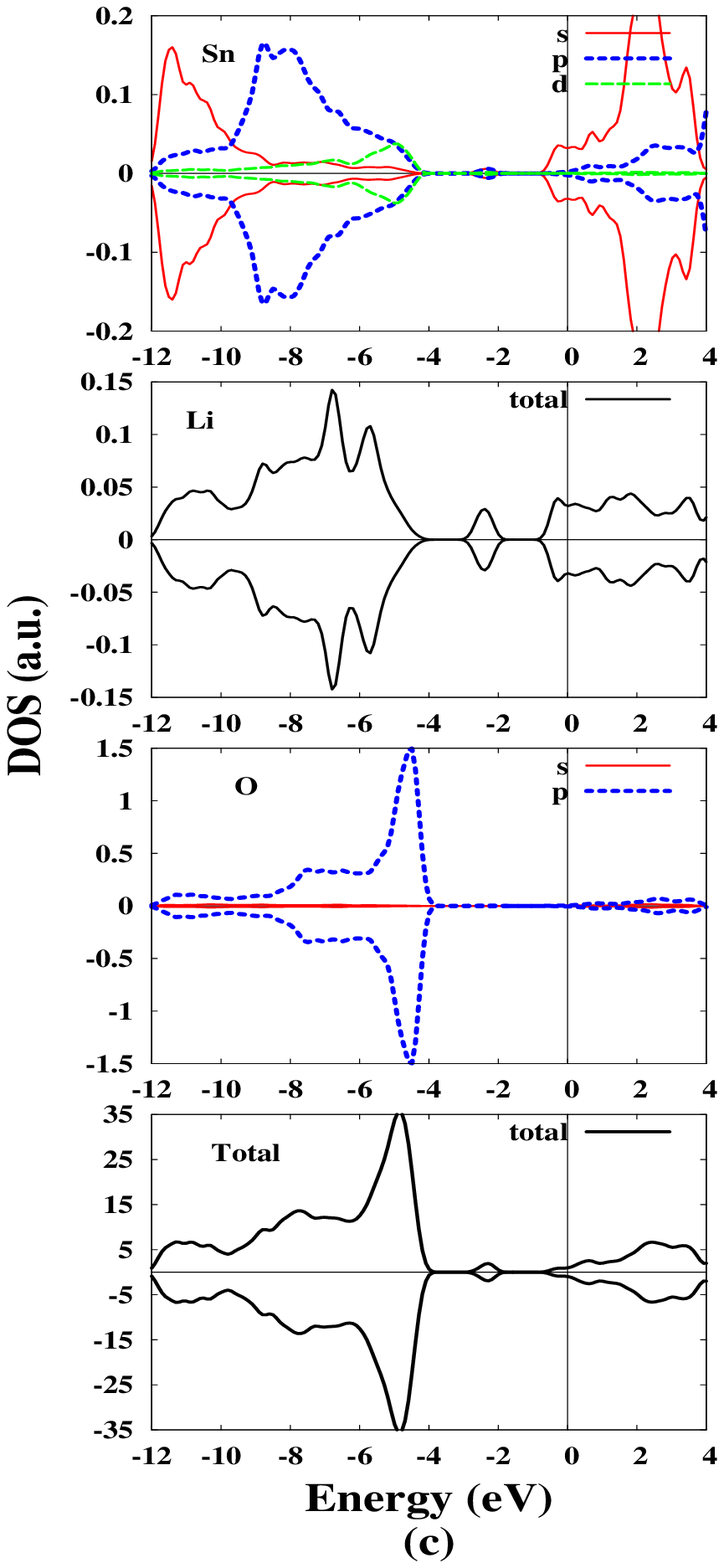,width=0.36\linewidth,height=1.2\linewidth,clip=}
\end{tabular}
\caption{(color online) The LDA$+U$ calculated total and partial density of states (DOS) of (a) V$_\mathrm{O}$+Li$_\mathrm{Sn}$ (b) V$_\mathrm{O}$+Li$_\mathrm{O}$, and (c) V$_\mathrm{O}$+Li$_\mathrm{int}$ systems. Solid (red) and dashed (blue, green) lines represent $s$, $p$ and $d$ states, respectively. The bottom panel solid line represents the total DOS. The positive (negative) DOS shows majority (minority) spin states. The Fermi level ($E_\mathrm{F}$) is set to zero.}
\label{dos3}
\end{figure}

Before summarizing our work, the electronic structure of the above mentioned defect complexes will be presented. Figure~\ref{dos3}(a) shows that the LDA$+U$ electronic structure is similar to Li$_\mathrm{Sn}$, i.e., the magnetism is mainly contributed by the O $p$ orbitals and the Li-O type hybridization can also be seen near the Fermi level. Oxygen or tin  vacancies localize the oxygen $p$ orbitals of the other O atoms in the presence of Li$_\mathrm{Sn}$ (see Fig.~\ref{dos2}(a) and Fig.~\ref{dos3}(a)). No magnetism can be seen in Fig.~\ref{dos3}(b), which represents the V$_\mathrm{O}$+Li$_\mathrm{O}$ defect system. The impurity peak around 2~eV below $E_\mathrm{F}$ is contributed by the Li atom. The conduction and valance bands are separated from each other and the Fermi level crosses the conduction band, which is mainly formed by Sn $s$ and Li orbitals. This electronic structure indicates that Li in the V$_\mathrm{O}$ system will behave as an electron donor. The electronic structure of V$_\mathrm{O}$+Li$_\mathrm{int}$ [Fig.~\ref{dos3}(c)] also shows nonmagnetic behavior. The O $p$ orbitals are completely occupied and are far from $E_\mathrm{F}$. The Li and Sn atoms occupy the conduction band, which is also crossed by the Fermi level, i.e.  V$_\mathrm{O}$+Li$_\mathrm{int}$ is also a $n$-type material. Comparing the DOS of the Li$_\mathrm{int}$ system with that of the V$_\mathrm{O}$+Li$_\mathrm{int}$, we see that V$_\mathrm{O}$ pushes the valence band of Li$_\mathrm{int}$ to lower energies. From the given DOS in Figs. (\ref{dos1}), (\ref{dos2}), and (\ref{dos3}), we may say that those systems in which $E_\mathrm{F}$ cuts the conduction bands, i.e., which behave as electron donors, do not show magnetism. This indicates that magnetism in SnO$_{2}$ is mediated by holes and destroyed by electrons. Further experimental work is needed to validate these theoretical predictions.

We have shown that the band gap of SnO$_{2}$ can easily be recovered using the LDA$+U$ approach. It is also interesting to note that the defect formation energies of defective systems can be improved by LDA$+U$, according to calculations on binding energies of similar defective systems.\cite{ref1,ref1a,ref10}   However, recent progress in DFT shows that defect states can be more accurately described  with hybrid functionals, such as the (atomic) self-interaction corrected functional (SIC or ASIC) or the Heyd, Scuseria, and Ernzerhof (HSE) functional. \cite{ref2,ref3, ref4, ref7, ref8, ref9, ref10}  These approaches can accurately account for the self interaction correction, which not only act upon the levels on which $U$ is applied but on the whole electronic structure. Hence, more accurate defect states are expected from such type of calculations. However, this does not change the main conclusion of our work, as similarly pointed out by Janotti \textit{et al}.,\cite{ref10} who compared the defect formation energies of ZnO using LDA$+U$ and hybrid approaches. Carter \textit{et al.}\cite{ref4} also studied defects in GaN using GGA and SIC, and both approaches gave qualitatively the same results. A recent comparative study between the LDA+ U and the HSE hybrid functional in ZnO shows that the defects thermodynamics can correctly be described by LDA$+U.~$\cite{ref2} Therefore, although it is expected that SIC (or ASIC) and HSE would have more profound effects on the electronic structure, we believe that LDA$+U$ can correctly describe defects in SnO$_{2}$, at least qualitatively.

A note of caution should be added however when comparing the formation energies of oxygen defects with those of other defects, since the application of the $U$ functional on the oxygen $p$ states particularly improves the electronic structure of oxygen related defects, as opposed to a bare-LDA less accurate treatment of other cases. However, the relative trends with/without $U$ are the same for all cases and the differences induced by the $U$ on the oxygen defects are comparable to the rest of defects (although larger, see Table 1). For these reasons we believe these results are qualitatively correct. Notice again that a more general treatment with hybrid functionals such as SIC ~\cite{ref8, ref4} or HSE ~\cite{ref3} would affect more democratically the whole electronic structure and improve the results.

\section{summary}

We investigated the energetics and magnetism of Li-doped SnO$_{2}$ systems with and without native defects using density functional theory (DFT) with LDA and LDA$+U$. Lithium was doped at Sn, O, and interstitial sites and it was shown that it can induce magnetism in SnO$_{2}$ when doped at the Sn site. No magnetism was found however when Li was doped at the O and interstitial sites. The defect formation energies showed that Li doped at the Sn site is more favorable than doped at the O site. We found that Li at the Sn site also shows a large magnetic moment ($3.00\mu_\mathrm{{B}}$), and the origin of magnetism was discussed in terms of electronic structures and spin densities. Native defects, Sn vacancies (V$_\mathrm{Sn}$) and O vacancies (V$_\mathrm{O}$), were also studied and it was observed that magnetic V$_\mathrm{Sn}$ has higher defect formation energy than non magnetic V$_\mathrm{O}$. To reduce the defect formation energies of these native defects, the interactions of Li with V$_\mathrm{Sn}$ and V$_\mathrm{O}$ were also studied. The calculated defect formation energies of these native defects were significantly decreased by Li. Our calculations showed that  V$_\mathrm{O}$ helps to promote magnetism in Li doped  SnO$_{2}$, in agreement with general experimental speculations. Structural relaxations were shown to be important when the interactions of Li with V$_\mathrm{Sn}$ were considered. Comparison of our studied system with previously synthesized systems was also discussed and it was concluded that Li doped SnO$_{2}$ is another good candidate in the filed of spintronics.

\acknowledgments
We are grateful to S. K. Hasanain for useful discussions on the experimental realization of the studied systems.
GR acknowledges the cluster facilities of NCP, Pakistan.
VMGS thanks the Spanish Ministerio de Econom\'{\i}a y Competitividad for a Ram\'on y Cajal fellowship (RYC-2010-06053). We acknowledge the use of computer resources (Altamira node) from the Spanish Supercomputing Network (RES).

\clearpage
\begin{table}
\caption{Formation energies calculated under equilibrium ($E_\mathrm{eq}$), Sn-rich ($E_\mathrm{Sn}$) and O-rich ($E_\mathrm{O}$) conditions, in units (eV). The last column lists magnetic moments (MM) per supercell, calculated in units of $\mu_\mathrm{B}$. Values in parentheses show formation energies calculated with LDA$+U$.}
\begin{tabular}{ccccccccc}
\toprule
System &$E_\mathrm{eq}$  &$E_\mathrm{Sn}$  &$E_\mathrm{O}$ & MM\\
\colrule
Li$_\mathrm{Sn}$&\,-1.03(-0.84)&\,\, 6.53\,(\,6.14)&\,-1.03(-0.84)&3.00(3.00)\\
Li$_\mathrm{O}$&\,-0.85\,\,(\,\,1.83)&\,\,-0.85\,(1.83)&\,\,\,2.93\,(\,5.32)&0.00(0.00)\\
Li$_\mathrm{int}$&\,\,-2.58(\,-2.05)&\,-2.58(-2.05)&\,-2.58(-2.05)&0.00(0.00) \\
V$_\mathrm{Sn}$&\,\,\,6.87\,\,(\,8.64)&14.44(15.62)&\,\,\,6.87\,(\,8.64)&4.00(4.00) \\
V$_\mathrm{O}$&\,\,\,1.64\,\,(\,4.40)&\,\,\,1.64\,\,(\,4.40)&\,\,\,5.42\,(\,7.89)&0.00(0.00)\\
\botrule
\end{tabular}
\label{t1}
\end{table}

\begin{table}
\caption{Formation energies (in units of eV) of systems  V$_\mathrm{Sn}$ + Li$_\mathrm{Sn}$, V$_\mathrm{Sn}$ + Li$_\mathrm{O}$, and V$_\mathrm{Sn}$ + Li$_\mathrm{int}$ calculated under equilibrium ($E_\mathrm{eq}$), Sn-rich ($E_\mathrm{Sn}$) and O-rich ($E_\mathrm{O}$) conditions. The distance from V$_\mathrm{Sn}$ to Li$_\mathrm{Sn}$, Li$_\mathrm{O}$, and Li$_\mathrm{int}$ is $r-r_{j}$ (in units of \AA). The last column lists the calculated magnetic moments (MM) per supercell (in units of $\mu_\mathrm{B}$). Values in parentheses show formation energies calculated with LDA$+U$.}
\begin{tabular}{cccccccccccc}
\toprule
System& $r-r_{j}$ &$E_\mathrm{eq}$ &$E_\mathrm{Sn}$&$E_\mathrm{O}$& MM \\
\colrule
V$_\mathrm{Sn}$ + Li$_\mathrm{Sn_j}$\\\cline{1-1}
V$_\mathrm{Sn}$ + Li$_\mathrm{Sn1}$&3.23&\,\,1.55(1.77)&\,\,\,9.11(8.75)&1.55(1.77)&7.00(7.00) \\
V$_\mathrm{Sn}$ + Li$_\mathrm{Sn2}$&3.76&\,\,2.13(2.64)&\,\,\,9.69(9.62)&2.13(2.64)&7.00(7.00)\\
V$_\mathrm{Sn}$ + Li$_\mathrm{Sn3}$&4.80&\,\,1.54(2.45)&\,\,\,9.10(9.43)&1.54(2.45)&7.00(7.00) \\
V$_\mathrm{Sn}$ + Li$_\mathrm{Sn4}$&5.78&\,\,1.56(2.42)&\,\,\,9.13(9.40)&1.56(2.42)&7.00(7.00)\\
V$_\mathrm{Sn}$ + Li$_\mathrm{Sn5}$&6.79&\,\,2.80(2.65)&10.36(9.63)&2.80(2.65)&7.00(7.00)\\
V$_\mathrm{Sn}$ + Li$_\mathrm{Sn6}$&7.52&\,\,2.91(2.80)&10.48(9.78)&2.91(2.80)&7.00(7.00) \\
V$_\mathrm{Sn}$ + Li$_\mathrm{Oj}$\\\cline{1-1}
V$_\mathrm{Sn}$ + Li$_{\mathrm{OI}}$&2.04&-4.16(\,-2.58)&\,\,\,3.40(4.40)&-0.38(\,0.91)&1.00(1.00)\\
V$_\mathrm{Sn}$ + Li$_{\mathrm{OII}}$&2.08&-1.74(\,-1.84)&\,\,\,3.82(5.14)&\,\,0.04(\,1.65)&1.00(1.00)\\
V$_\mathrm{Sn}$ + Li$_{\mathrm{OIII}}$&4.71&\,\,6.09(12.56)&13.65(19.54)&\,\,4.93(16.05)&1.00(1.00)\\
V$_\mathrm{Sn}$ + Li$_\mathrm{int}$\\\cline{1-1}
V$_\mathrm{Sn}$ + Li$_\mathrm{int}$&5.73&\,\,1.83(2.35)&\,\,9.40(9.32)&1.83(\,2.35)&3.00(3.00) \\
\botrule
\end{tabular}\\
\label{t2}
\end{table}

\clearpage
\begin{table}
\caption{Formation energies (in units of eV) of systems  V$_\mathrm{O}$ + Li$_\mathrm{Sn}$, V$_\mathrm{O}$ + Li$_\mathrm{O}$, and V$_\mathrm{O}$ + Li$_\mathrm{int}$ calculated under equilibrium ($E_\mathrm{eq}$), Sn-rich ($E_\mathrm{Sn}$) and O-rich ($E_\mathrm{O}$) conditions. The distance from V$_\mathrm{O}$ to Li$_\mathrm{Sn}$, Li$_\mathrm{O}$, and Li$_\mathrm{int}$ is $r-r_{j}$ (in units of \AA). The last column lists the calculated magnetic moments (MM) per supercell (in units of $\mu_\mathrm{B}$). Values in parentheses show formation energies calculated with LDA$+U$.}
\begin{tabular}{ccccccccccccccc}
\toprule
System &$r-r_{j}$&$E_\mathrm{eq}$ &$E_\mathrm{Sn}$ &$E_\mathrm{O}$& MM \\
\colrule
V$_\mathrm{Oj}$ + Li$_\mathrm{Sn}$\\\cline{1-1}
V$_\mathrm{O1}$ + Li$_\mathrm{Sn}$&2.04&-2.55(-1.88)&5.01(\,5.10)&\,\,1.23(1.61)&1.00(1.00)\\
V$_\mathrm{O2}$ + Li$_\mathrm{Sn}$&2.08&-2.34(-2.76)&5.22(\,4.22)&\,\,1.44(0.73)&1.00(1.00)\\
V$_\mathrm{O3}$ + Li$_\mathrm{Sn}$&4.71&-3.69(-0.02)&3.87(6.96)&\,\,0.09(3.47)&1.00(1.00)\\
V$_\mathrm{O4}$ + Li$_\mathrm{Sn}$&5.71&-2.09(0.05)&3.39(\,7.02)&-0.39(3.54)&1.00(1.00)\\
V$_\mathrm{O}$ + Li$_\mathrm{Oj}$\\\cline{1-1}
V$_\mathrm{O}$ + Li$_\mathrm{O1}$&2.94&0.15(2.57)&0.15(2.57)&3.93(6.06)&0.00(0.00)\\
V$_\mathrm{O}$ + Li$_\mathrm{O2}$&3.47&0.21(2.71)&0.21(2.71)&3.98(6.20)&0.00(0.00)\\
V$_\mathrm{O}$ + Li$_\mathrm{O3}$&5.15&0.30(3.06)&0.30(3.06)&4.08(6.55)&0.00(0.00)\\
V$_\mathrm{O}$ + Li$_\mathrm{int}$\\\cline{1-1}
V$_\mathrm{O}$ + Li$_\mathrm{int}$&3.48&-0.96(1.63)&-0.96(\,1.63)&1.36(5.11) & 0.00(0.00)\\
\botrule
\end{tabular}\\
\label{t3}
\end{table}
\clearpage

\clearpage

\clearpage

\clearpage

\clearpage


\begin{thebibliography}{99}

%
\bibitem{ZnO} K. Ueda, H.Tabata, and T.Kawai, Appl. Phys. Lett. \textbf{79}, 988 (2001).

\bibitem{ZO} K. Yang, R. Wu, L. Shen, Y. Ping Feng, Y. Dai, and B. Huang, Phys. Rev. B \textbf{81}, 125211 (2010).


\bibitem{tio2} K. S. Yang, Y. Dai, and B. B. Huang, J. Chem. Phys. \textbf{10}, 2327 (2009).

\bibitem{tio3} K. S. Yang, Y. Dai, B. B. Huang, and M. H. Whangbo, Appl. Phys. Lett.\textbf{93}, 132507 (2008).

\bibitem{tio4}K. S. Yang, Y. Dai, B. B. Huang, and Y. P. Feng, Phys. Rev. B \textbf{81}, 033202 (2010).

\bibitem{ogale} S. B. Ogale1, R. J. Choudhary, J. P. Buban, S. E. Lofland, S. R. Shinde, S. N. Kale, V. N. Kulkarni, J. Higgins, C. Lanci, J. R. Simpson, N. D. Browning, S. Das Sarma, H. D. Drew, R. L. Greene, and T. Venkatesan, Phys. Rev. Lett. \textbf{91}, 077205 (2003).

\bibitem{gu} G. Rahman, {V\'{\i}ctor M. Garc\'{\i}a-Su\'arez}, and S. C.  Hong, Phys. Rev. B \textbf{78}, 184404 (2008).

\bibitem{znoprl-2005}M. H. F. Sluiter, Y. Kawazoe, P. Sharma,A. Inoue, A. R. Raju, C. Rout, and U.V. Waghmare, Phys. Rev. Lett. \textbf{94}, 187204 (2005).


\bibitem{01} M. Khalid, M. Ziese, A. Setzer, P. Esquinazi, M. Lorenz, H. Hochmuth, M. Grundmann, D. Spemann, T. Butz, G. Brauer, W. Anwand, G. Fischer, W. A. Adeagbo, W. Hergert, and A. Ernst, Phy. Rev. B \textbf{80}, 035331 (2009).

\bibitem{02}D. Kim, J. Hong, Y. Ran Park and K. Joo Kim, J. Phys.: Condens. Matter \textbf{21}, 195405 (2009).

\bibitem{03}F. M\'{a}ca, J. Kudrnovsk\'{y}, V. Drchal, and Georges Bouzerar, Appl. Phys. Lett. \textbf{92}, 212503 (2008).

\bibitem{04} L. X. Guan, J. G. Tao, C. H. A. Huan, J. L. Kuo, and L. Wang, J. Appl. Phys. \textbf{108}, 093911 (2010).

\bibitem{05}V. Fernandes, P. Schio, A. J A de Oliveira, W. A. Ortiz, P. Fichtner, L. Amaral, I. L. Graff, J. Varalda, N. Mattoso, W.H. Schreiner and D. H. Mosca, J. Phys.: Condens. Matter \textbf{22}  216004 (2010).

\bibitem{06} J. Osorio-Guill\'{e}n, S. Lany, S. V. Barabash, and A. Zunger, Phy. Rev. B \textbf{75}, 184421 (2007).

\bibitem{KCa} S. K. Srivastava, P. Lejay, B. Barbara, S. Pailh\'{e}s, V. Madigou, and G. Bouzerar, Phys. Rev. B \textbf{82}, 193203 (2010).

\bibitem{Nit} W. Zhi Xiao, L. Ling Wang, L.  Xu, Q. Wan, and Zou, Solid State Communications \textbf{149}, 1304-1307 (2009).

\bibitem{Mg}C. Wen Zhang, and S. shen Yan, Appl. Phys. Lett. \textbf{95}, 232108 (2009).

\bibitem{Rahman2010} G. Rahman and V. M. Garc\'{i}a-Su\'{a}rez, Appl. Phys. Lett. \textbf{96}, 052508(2010).

\bibitem{Hong}N. H. Hong, J. H. Song, A. T. Raghavender, T. Asaeda, and M. Kurisu, Appl. Phys. Lett. \textbf{99}, 052505 (2011).

\bibitem{cexp} N. Hoa Hong, J. H. Song, A. T. Raghavender, T. Asaeda, and M. Kurisu, Apl. Phys. Lett \textbf{99}, 052505 (2011).

\bibitem{ZnOprl} J. B. Yi, C. C. Lim, G. Z. Xing, H. M. Fan, L. H. Van, S. L. Huang, K. S. Yang, X. L. Huang, X. B. Qin, B. Y. Wang, T. Wu, L. Wang, H. T. Zhang, X. Y. Gao, T. Liu, A. T. S. Wee, Y. P. Feng, and J. Ding, Phys. Rev. Lett. \textbf{104}, 137201 (2010).

\bibitem{S.Ghosh}S. Ghosh, G. G. Khan, and K. Mandal, Appl Mater Interfaces. \textbf{4}, 2048 (2012).

\bibitem{Chang} G. S. Chang, J. Forrest, E. Z. Kurmaev, A. N. Morozovska, M. D. Glinchuk, J. A. McLeod, A. Moewes, T. P. Surkova, and N. Hoa Hong, Phys. Rev. B \textbf{85}, 165319 (2012).

\bibitem{Vgolo} V. Golovanov, N. Ozcan, M. Viitala, T. T. Rantala and J. Vaara, \textit{NATO Science for Peace and Security Series B: Physics and Biophysics, 2012, Part 2, 315}

\bibitem{Wang} C. Wang, Q. Wu, H. L. Ge, T. Shang, and J. Z. Jiang Nanotechnology \textbf{23}, 075704(2012).

\bibitem{alex} C. Klyc and A. Zunger, Phys. Rev. Lett. \textbf{88}, 095501 (2002).

\bibitem{stefano} C. D. Pemmaraju and S. Sanvito, Phys. Rev. Lett.
\textbf{94}, 217205(2005).

\bibitem{Zn}W. Wei, Y. Dai, M. Guo, K. Lai, and B. Huang, J. Appl. Phys. \textbf{108}, 093901 (2010).

\bibitem{chro}W. Wei,Y. Dai, M. Guo,Z. Zhang,and B. Huang, J. Solid State Chem \textbf{183}, 3073-3077 (2010).

\bibitem{ptype}M. Mehdi, B. Mohagheghi, and M. Shokooh-Saremi, Semicond. Sci. Technol. \textbf{19},  764 (2004).

\bibitem{group}J. Haines and J. M. L\'{e}ger, Phys. Rev. B \textbf{55}, 11144 (1997), A. A. Bolzan, C. Fong, B. J. Kennedy, and C. J. Howard, Acta Crystallogr. B \textbf{53}, 373 (1997).


\bibitem{DFT} P. Hohenberg and W. Kohn, Phys. Rev. \textbf{136}, B846 (1964).

\bibitem{siesta} J. M. Soler et al J. Phys.Condens.: Matter \textbf{14}, 2745 (2002).

\bibitem{lda} J. P. Perdew  and A. Zunger, Phys. Rev. B \textbf{23}, 5048 (1981).

\bibitem{ps} D. R. Hamann , M. Schl\'{u}ter  and C. Chiang,   Phys. Rev. Lett. \textbf{43}, 1494 (1979).

\bibitem{pss}L. Kleinman  and D. M. Bylander, Phys. Rev. Lett. \textbf{48}, 1425 (1982).

\bibitem{cg}W. H. Press, B. P. Flannery, S. A. Teukolsky, W. T. Vetterling, New Numerical Recipes, Cambridge University Press, New York, 1986.

\bibitem{ldau1}I. Nekrasov, M. Korotin, and V. Anisimov, arXiv:
cond-mat/0009107v1.

\bibitem{ldau2} S.-G. Park, B. M.-K\"{o}pe,and Y. Nishi, Phys. Rev. B \textbf{82}, 115109 (2010).

\bibitem{VASPc}G. Kresse and J. Furthmüller, Phys. Rev. B, {\bf 54}, 11169 (1996).

\bibitem{defects-2012} K. Yang, Y. Dai, and B. Huang, arXiv:
cond-mat/1202.5651v1


\bibitem{form} A. F. Kohan, G. Ceder, D. Morgan, and C. G. V. de Walle,  Phys. Rev. B \textbf{61}, 15019 (2000).

\bibitem{PRB2012}J. A. Berger,L. Reining, and F. Sottile, Phys. Rev. B \textbf{85}, 085126 (2012).

\bibitem{iop} Silva \textit{et. al}, Physica Scripta, \textbf{T109}, 180 (2004).

\bibitem{dean}J. A. Dean (ed.), Lange's Handbook of Chemistry, fourteenth
ed. (McGraw-Hill, Inc., New York, 1992).

\bibitem{Sanvito} A. Droghetti, C. D. Pemmaraju, and S. Sanvito, Phys. Rev. B \textbf{78}, 140404 R (2008).
\bibitem{Fernandes} V. Fernandes, R. J. O. Mossanek, P. Schio, J. J. Klein, A. J. A. de Oliveira, W. A. Ortiz, N. Mattoso, J. Varalda,
W. H. Schreiner, M. Abbate, and D. H. Mosca, Phys. Rev. B \textbf{80}, 035202 (2009).

\bibitem{Rahman2013} {G. Rahman}, {V\'{\i}ctor M. Garc\'{\i}a-Su\'arez}, and J. M. Morbec, J. Mag. Mag. Mater.  \textbf{328}, 104(2013).


\bibitem{juliana2013} {J. M. Morbec and G. Rahman}   arXiv: cond-mat/1207.6342v3

\bibitem{SnO2Li-2013} While revising our manuscript, we found that Srivastava  \textit{et al.} {arXiv: cond-mat/1302.4869v1 }  have found  ferromagnetism in Li doped SnO$_{2}$. Their results confirmed our theoretical work.     

\bibitem{gul2}J. M. Themlin, R. Sporken,  J. Darville, R. Caudano,  and J. M. Gilles, Phys.Rev. B \textbf{42}, 11915 (1990).

\bibitem{refs-exp2} C. Zhang and S. Yan, J. Appl. Phys. \textbf{106}, 063709 (2009).

\bibitem{Nidoped} H. Wang, Y. Yan, X. Du,  X. Liu, K. Li, and H. Jin, J. Appl. Phys. \textbf{107}, 103923  (2010).

\bibitem{refs-exp} N. H. Hong, J. Sakai, W. Prellier, and A. Hassini, J. Phys.: Condens.Matter \textbf{17}, 1697 (2005).

\bibitem{ref1} P. Reunchan , X. Zhou , S. Limpijumnong , A. Janotti , and C. G. V. de Walle , Current Applied Physics \textbf{11}, S296eS300(2011) 


\bibitem{ref1a}
S.-G. Park, B. Magyari-K\''ope, and Y. Nishi, Phys. Rev. B \textbf{82}, 115109 (2010).


\bibitem{ref10}
A. Janotti and C. G. Van de Walle, phys. status solidi B \textbf{248}, No. 4, 799 (2011).

\bibitem{ref2} R. Sanizn, Y. Xu , M. Matsubara, M.N. Amini, H. Dixit, D. Lamoen, and B. Partoens,  J. Phys. Chem. Solids \textbf{74}, 45 (2013).



\bibitem{ref3}
A. Janotti,J. B. Varley, P. Rinke, N. Umezawa, G. Kresse, and C. G. Van de Walle, Phys. Rev B \textbf{81}, 085212 (2010).


\bibitem{ref4}

D.J Carter , M. Fuchs and C. Stampf, J. Phys.: Condens. Matter \textbf{24} 255801 (2012).
\bibitem{ref7}
R. Ramprasad, H. Zhu, P. Rinke, and M. Scheffler, Phys. Rev Lett. \textbf{108}, 066404 (2012).

\bibitem{ref8}
C. D. Pemmaraju, T. Archer, D. S\'anchez-Portal, and S. Sanvito, 
Phys. Rev B \textbf{75}, 045101 (2007).



\bibitem{ref9} P. S\'piewak, J. Vanhellemont, and K. J. Kurzydlowski , J. Appl. Phys. \textbf{110}, 063534 (2011).


\end{thebibliography}
\end{document}